\documentstyle[11pt,newpasp,twoside]{article}
\begin{document}
\title{Massive clusters in Hubble sequence galaxies: Ellipticals}
 \author{Markus Kissler-Patig}
\affil{European Southern Observatory, Karl-Schwarzschild-Str.~2, D-85748
Garching, Germany}

\begin{abstract}
A brief overview of the properties of massive star clusters in early-type
galaxies is given. All ellipticals (with only one known exception) host
massive star clusters in the form of globular clusters, suggesting that
their formation is very common. The number of clusters per unit galaxy mass 
appears very constant, pointing at similar formation processes on large
scales. Furthermore, the luminosity/mass functions of the globular cluster 
systems appear to be very similar from galaxy to galaxy, pointing at similar 
formation processes on small scales.   

Probably the most important recent discovery is the presence of cluster 
sub-populations in many (if not all) early-type galaxies. The latter seem to be
closely linked to galaxy formation and evolution, and appear to support
hierarchical clustering models, however with an early spheroid formation. 
\end{abstract}


\section{Introduction}

This short contribution is aimed at giving an {\it overview} of the
current state of the research on globular clusters systems in early-type
galaxies. It is, obviously, not aimed at giving a {\it review} on the
subject. Excellent reviews have recently been written by Ashman \& Zepf (1998),
and Harris (1999a,b). I will focus on aspects related to the formation of
massive clusters rather than on what can be learned from similar studies
about the formation and evolution of the host galaxies.
In early-type galaxies showing no signs of a recent star formation
event, massive star clusters are exclusively globular clusters.

After a brief historical introduction (Sect.~2), similarities of the globular
cluster properties in all ellipticals are highlighted (Sect.~3).
Section 4 addresses the important aspect of multiple globular
cluster sub-populations in ellipticals, before a brief summary is
presented in Sect.~5.

\section{A brief history}

Globular clusters in spirals (and in particular in our Milky Way) were studied 
for a century already. Shapley (1918) was the first to consider the
system of globular clusters as en entity in our Milky Way. Hubble (1932)
followed up with the globular cluster system of M31. But it lasted
another 23 years before Baum (1955) suggested the presence of globular clusters
in an elliptical (M~87), and another 13 years before Racine (1968) studied
the colors and luminosities of these clusters. The first thesis on
extragalactic globular cluster systems was composed by Hanes (1977) and 
launched what is now a very active field of research. The 80's saw a number of 
studies with the first generation of CCDs. A number of properties of globular
cluster systems were identified and correlations with host-galaxy
properties observed (see Harris 1991 for a review of the 80's). The
90's, with a new generation of wide field imagers and the WFPCs on
board of HST, allowed a number of more detailed studies for $\sim 100$
galaxies. One of the most important discoveries of the last decade was
the presence of several globular cluster sub-populations around many
early-type galaxies (Zepf \& Ashman 1993, Gebhardt \& Kissler-Patig
1999), which will be commented in Sect.~4.
Thus, globular clusters in spirals are know for a century already, 
but globular clusters in ellipticals have a much shorter history,
which should be kept in mind when assessing the conclusions 
draw from a limited number of studies. 

\section{General properties of globular cluster systems and what they
tell us about the formation of massive clusters}

\subsection{Their presence around all galaxies}

As emphasized already by Harris (1991), all early-type galaxies host
globular clusters. Studies since then only confirmed this trend in the
sense that no galaxy was observed not to host any globular clusters. The
number of observed clusters in normal (non-dwarf) galaxies ranges from a
few hundreds to a few thousands.

The only exception to date is M~32, the low-luminosity ($M_V\sim-16.4$),
high surface-brightness companion of M~31, in which 15-20 globular clusters 
are expected, but none was detected. It remains unclear whether its population
could have been efficiently removed during interactions with M~31, or
destroyed by dynamical friction and tidal shocks from the dense nucleus
of M~32 itself. M~32 remains a minor puzzle.

Nevertheless, the other $\sim100$ cases demonstrate that the formation
of massive star clusters during the formation and evolution of
early-type galaxies is very common. The formation of star clusters as a common
mode of star formation might not be surprising, but the abundant presence
of globular clusters in all early-type galaxies is one of its best proofs.

\subsection{Their number per unit luminosity}

To compare the relative numbers of globular clusters in galaxies of
different sizes, Harris \& van den Bergh (1981) introduced the specific
frequency: the number of globular clusters per unit (M$_V=10^{15}$) of
galaxy light. The specific frequency of early-type galaxies was found to
be pretty constant except in central cD galaxies which appeared to exhibited a
large overabundance of globular clusters. The latter galaxies, however, were 
shown not to be overabundant in globular clusters when the total mass
(in particular the hot gas) surrounding them was properly taken into account 
(see McLaughlin 1999, and in these proceedings).

To better compare spirals and
ellipticals, Zepf \& Ashman (1993) introduce a quantity similar to the
specific frequency, but normalized to mass (taking into account the different
mass-to-light ratio as a function of galaxy type). Elliptical galaxies
then showed a marginal overabundance when compared to spirals, 
suggesting that globular
clusters formed more efficiently with respect to stars during the formation
of spheroids than
during the formation of the halo (whose clusters dominate late-type
galaxies). Note, however, that McLaughlin (1999) showed that the 
halo/spheroid of the Milky Way (excluding the disk) had a similar globular 
cluster formation efficiency as ellipitcals when normalized to the total 
baryonic mass, which emphasizes the universal efficiency for cluster formation.

Thus, not only is the formation of globular clusters common, but it appears
to occur with a roughly constant efficiency independent of the galaxy type,
environment or metallicity.

\subsection{Their luminosity/mass function}

Another ``universal'' property of globular clusters is their luminosity
or mass function that appears to have a characteristic scale (see also
the contribution of Miller; Fritze-von Alvensleben; and McLaughlin in
these proceedings). 

The constancy of the turn-over magnitude of the globular cluster
luminosity function was recognized early on and explored as a standard
candle (cf.~Hanes 1979). Various authors claimed that a dependence of
galaxy type or environment could exist (e.g.~Whitmore 1997 for a summary). 
It appears, however, very likely that the underlying mass distributions 
are very similar and that the slightly different turn-overs in the luminosity
functions can be fully explained by the dependency of
the luminosity from age and metallicity. Towards the center of galaxies
(closer than $\sim5$ kpc) dynamical effects will obviously also play a role.

Various arguments lead to believe that the characteristic mass in the
mass distributions of globular clusters is not a product of dynamical
evolution of the system (cf.~the above mentioned contributions to these
proceedings). One of the arguments is the presence of this same
characteristic mass in galaxies with very different gravitational
potential including dwarf galaxies, spirals, cuspy and core ellipticals
of various masses. It rather seems that the characteristic scale is
implemented right from the beginning, although it remains unclear
whether this is linked to the exact formation process of massive
clusters inside a given molecular cloud, or whether this is an
implication of a characteristic scale already present in the mass
distribution of the molecular clouds at the origin of the globular
clusters. Both cases, however, support a ``universal'' and very
homogeneous formation process of globular clusters in the universe, also
on the small scales of molecular clouds.

\section{The presence of multiple globular cluster sub-populations in
early-type galaxies}

\subsection{Discovery and frequency of multiple globular cluster
populations in early-type galaxies}

One of the currently most interesting aspects of globular cluster systems of
early-type galaxies is the presence of multiple sub-populations in many
galaxies. In retrospective, this discovery could have been anticipated
since the discovery of halo and bulge clusters in the Milky Way
(e.g.~Jablonka in the proceedings). It came, nevertheless, as a surprise
when these sub-populations were first discovered in globular
cluster color distributions by Zepf \& Ashman (1993). The most recent
studies suggest that $\sim 50$ \% of all early-type galaxies host more
than one globular cluster population (Gebhardt \& Kissler-Patig 1999,
see also Kundu 1999). This number is only a lower limit given the fact that the
studies were conducted with $V-I$ colors that are not very sensitive to
metallicity. The absence of multi-modal distributions around more
galaxies could therefore be due to the observational limits combined
with  the age-metallicity degeneracy present in optical colors, rather than
to the physical absence of multiple sub-populations in these galaxies
(see also Kissler-Patig, Forbes, Minniti 1998).

\subsection{The properties of metal-rich and metal-poor populations}

The reality of the distinction, first observed in colors only, was established 
in studies that derived very different spatial distributions for the metal-poor
and metal-rich clusters around their host galaxies (to date, NGC 1380:
Kissler-Patig et al.~1997; NGC 4472: Lee et al.~1998; NGC 3115: Kundu \&
Whitmore 1998).

Generally speaking, the metal-poor clusters have a more spherical (or
less flattened) distribution and are somewhat more extended than the diffuse 
stellar light, while the metal-rich population closely follows the ellipticity
and position angle of the observed (high surface-brightness) stellar light.
Thus, the metal-poor population gets associated with the halo,
while the red population appears associated with the spheroid/thick-disk
component. Interestingly, both from spectroscopic studies, as well as
from photometric studies, the metal-poor and metal-rich globular clusters
appear coeval and old, with ages similar to the ones of the Milky Way globular
clusters (see Kissler-Patig in these proceedings).

From dynamical studies, the metal-poor clusters in
ellipticals seem to dominate the rotation or at least to be on
tangentially biased orbits and have a high velocity dispersion, as opposed to 
the metal-rich clusters that exhibit little rotation, are  apparently on more
radial orbits and have a lower velocity dispersion (cf.~Kissler-Patig \&
Gebhardt 1998; Sharples et al.~1998; Kissler-Patig et al. 1999). 

Finally, the three studies in this respect (Kundu \& Whitmore 1998;
Puzia et al.~1999; Kundu et al.~1999) showed that the red globular clusters
are systematically smaller than the blue ones, and this at all radii out
to several kpc (tested up to $\sim 20$ kpc). Dynamical effects could play a 
role, given that the red clusters appear to be on more radially biased orbits,
however the reason for this discrepancy is still unclear.

\subsection{The origin of the sub-populations}

Deriving the origin of these sub-populations from their properties is
currently the subject of many papers. Originally, the presence of the
two populations was predicted by Ashman \& Zepf (1992) in the frame of
elliptical formation through spiral--spiral mergers. Their first simple
scenario turned out to have problems explaining all the properties of
the blue and red sub-populations (e.g.~Kissler-Patig et al.~1997; 
Forbes et al.~1997), and it became clear that several other scenarios
can explain the presence of the two populations (see Kissler-Patig 1997;
Kissler-Patig et al.~1998; C\^ot\'e et al.~1998; Harris et al.~1998; Hilker et
al.~1999; Harris et al.~1999). Which mode of globular cluster formation
dominates the building up of most globular cluster systems around
galaxies is the matter of current studies. The main questions being {\it
i)} whether or not the clusters associated with the bulge (and the
bulges themselves) primarily formed in gas-rich mergers, as opposed to
early-type collapses -- falling back on the older debate on galaxy
formation; {\it ii)} whether or not the metal-poor clusters formed
in association with their final host galaxies. The latter does not seem to be
the case (see Burgarella et al.~2000, and in these proceedings),
although it appears difficult to distinguish between a formation
completely decoupled from the galaxy and a formation in individual
fragments within the initial dark halo.

\section{Summary}

The common presence of globular clusters around early-type galaxies, as
well as their universal properties suggest a very homogeneous formation
of star clusters in the universe. The characteristic mass scale is
either telling us about their formation process or implies a
characteristic mass already present in the molecular clouds out of which
these clusters form.

The presence of halo and bulge clusters appears to be common. The origin
of these sub-population is still unclear. However, there is strong
evidence that the metal-poor clusters formed in fragments independently of 
their final host galaxies, supporting hierarchical clustering
models. The formation of the metal-rich clusters appears to be
associated with the formation of the spheroid (eventually thick disk),
but must have happened at early epochs given the old measured ages of
the metal-rich clusters.

\acknowledgments

I would like to thank the organizers and in particular Ariane Lan\c{c}on
for initiating and organizing such a pleasant meeting.

\section*{Discussion}

\noindent {\bf Daniella Calzetti:} You mentioned that the observed properties
of globular cluster populations do not support merging scenarios as only
process for galaxy formation. Can you clarify which are the problems with the
merging scenarios? 

\noindent {\bf Markus Kissler-Patig:} This was partly a statement to act
against the ``fashionable'', but in my opinion wrong, idea that the metal-rich
populations are the sole product of spiral--spiral mergers. I would like
to remind people that alternative scenarios exist which are perfectly
compatible with the data (mainly because no exact/unique predictions
exist for any scenario to date). Forbes et al.~(1997), for example, list a
number of problems disqualifying the very simple scenario as originally
proposed by Ashman \& Zepf (1992), mainly based on the ratios of blue to red 
objects. Furthermore, the old ages of the
metal-rich clusters in the studied system (in galaxy clusters only, so
far) rule out mergers since $z=1$, which caused trouble to hierarchical
merging scenarios when these still predicted the majority of
star-formation to happen between $z=0.5$ and 1.0. Similar arguments can
be made for other galaxies in which no sub-populations were detected yet
with the current observations (see Kissler-Patig, Forbes, \& Minniti 1998).
Finally, it remains to be shown that large red populations with the
observed properties (out to several tens of kpc) can be build up in
spiral--spiral mergers. In other words, whether the new clusters
observed in interacting systems have indeed globular properties
(individually and as a system).

\noindent {\bf Hans Zinnecker:} On the two populations of globular clusters
(metal-poor and metal-rich): could they have their origin in two
different types of galaxies merging, namely metal-poor Magellanic type
dwarf galaxies coming together early on (halo clusters) and somewhat
later a major collision of two more metal-rich spirals (bulge clusters)?
Have people working on hierarchical galaxy formation considered this?

\noindent {\bf Markus Kissler-Patig:} This is close to the most believed
scenarios to date, although the first merging you mentioned is often
referred to as accretion, and the accreted dwarf galaxies do not need to
be of Magellanic type (the whole is then not too far from an Searle \&
Zinn picture, except with bigger fragments and outside a well defined
initial halo). For the formation of the bulges, indeed a spiral--spiral
merger could be the case, but if I understand you right, the spirals
would not have bulges themselves, and remember that given the ages of
the bulge clusters this must have happened a very high redshift: we fall
back onto a formation of the bulge by one or more gas-rich entities.
Concerning hierarchical galaxy formation models: they handle mostly dark
halos, and the baryonic parts of these are usually viewed as disks of various
sizes. It is non-trivial to link these simulations with such exact terms as
Magellanic type dwarfs or spirals of a given type.

\noindent {\bf Pavel Kroupa:} Will a large number of mergers of gas-rich
galaxies not produce a Gaussian distribution of globular cluster
properties, assuming globular clusters form during mergers.

\noindent {\bf Markus Kissler-Patig:} Indeed a large number of globular
clusters formed in mergers will ``smear out'' the color distribution {\it
but only for the red (metal-rich) objects}. The metal-poor clusters are not
though to have formed in gas rich mergers, and C\^ot\'e et al.~(1998)
showed in simulations that for a steep galaxy luminosity function, a seed
galaxy would accrete a large number of low-luminosity dwarfs, building
up a distinct population of blue clusters, before a larger merger event
(from the high-luminosity end of the galaxy luminosity function) would
add a distinct metal-rich population that could later be smeared out by
further mergers. So
one should keep in mind that the metal-rich population is likely to be a
composite of several episodes. 

\noindent {\bf Georges Meylan:} Photometric observations are an
essential first step towards the study of globular cluster systems. But
large pollution is present: checks by radial velocities, in NGC 1399 and
NGC 1316, show that $\sim$ 50\% of the globular cluster candidates are actually
foreground stars and background galaxies. Consequently, spectroscopic
check is an essential second step.

\noindent {\bf Markus Kissler-Patig:} The studies you refer too were
conducted with the NTT, and prepared with ground-based, optical
photometry. These samples are partly artificially contaminated because
the masked were ``filled'' with bright objects when no good globular
cluster candidate was present. More recent studies, prepared with combinations
of optical and near-infrared photometry, as well as HST imaging, have
typical contaminations of less than 10\%--20\% (also because
10m-class telescopes allow the spectroscopy of fainter objects and do not
limit the choice of targets).  But I agree that the next large step
forward in the field will be spectroscopy for a very large number of
globular clusters with instruments such as VIMOS (VLT) or DEIMOS (Keck).
Also you are right that older ground-based photometric studies dealing with 
a limited number of globular clusters can be severally affected by
background contamination.

\noindent {\bf Torsten B\"oker:} Why is the bimodal distribution
attributed to metallicity effects alone, rather than e.g~age or
extinction?

\noindent {\bf Markus Kissler-Patig:} In the particular case of early-type
galaxies, internal extinction is not a major concern given the lack of
a significant amount of dust in these galaxies, at least at the large
radii at which the clusters are observed.
Further, the colors of old globular clusters are completely dominated by 
metallicity rather than by age. This was confirmed by all
spectroscopic surveys conducted to date. For example, a metallicity 
difference of 1 dex in [Fe/H] would have the same effect on $V-I$ as an
age difference of roughly 10 Gyr.

\end{document}